\documentclass{PoS}

\usepackage{amssymb}

\usepackage{amsmath}
\usepackage{bm}
\usepackage{slashed}

\usepackage{mciteplus}

\newlength{\dummysp}
\newcommand{\beq}{\begin{eqnarray}}
\newcommand{\eeq}{\end{eqnarray}}

\newcommand{\gappeq}{\mathrel{\rlap {\raise.5ex\hbox{$>$}}
{\lower.5ex\hbox{$\sim$}}}}
\newcommand{\lappeq}{\mathrel{\rlap{\raise.5ex\hbox{$<$}}
{\lower.5ex\hbox{$\sim$}}}}

\newcommand{\ben}{\begin{enumerate}}
\newcommand{\een}{\end{enumerate}}

\newcommand{\bit}{\begin{itemize}}
\newcommand{\eit}{\end{itemize}}

\newcommand{\avg}[1]{\left\langle#1\right\rangle}

 \newcommand{\be}{\begin{equation}}
 \newcommand{\ee}{\end{equation}}
 
\def\[{\left [}
\def\]{\right ]}
\def\({\left (}
\def\){\right )}

\title{Finite density $\mathbf{O(3)}$ non-linear sigma model and low energy physics}

\ShortTitle{Finite density $O(3)$ non-linear sigma model}
	
\author{Falk Bruckmann\footnote{Supported by the DFG (BR 2872/6-1).}\\
        Universit\"at Regensburg, Institut f\"ur Physik, Universit\"atstra{\ss}e 31, 93053 Regensburg, Germany\\
        \email{falk.bruckmann@physik.uni-regensburg.de}}

\author{Christof Gattringer, \speaker{Thomas Kloiber}\footnote{Supported by the Austrian Science Fund, FWF, DK {\sl Hadrons in Vacuum, Nuclei, and Stars} (FWF DK W1203).}\\
        Universit\"at Graz, Institut f\"ur Physik, Universit\"atsplatz 5, 8010 Graz, Austria\\
        \email{christof.gattringer@uni-graz.at}, \email{thomas.kloiber@uni-graz.at}}

\author{Tin Sulejmanpasic\\
        North Carolina State University, Department of Physics, Raleigh, NC 27695-8202, USA\\
        \email{tsulejm@ncsu.edu}}
        
\abstract{We present lattice results for simulations of the $O(3)$ non-linear sigma model at finite chemical potential. The complex action
 problem is overcome by a dual variable representation of the model. We discuss two aspects of the theory at finite density: 
 1) The relation of the finite density data to scattering phases and wave-functions. 2) The phase-structure of the theory as a
 function of chemical potential and the possibility of a Kosterlitz-Thouless phase transition.}

\FullConference{The 33rd International Symposium on Lattice Field Theory\\
		14 -18 July 2015\\
		Kobe International Conference Center, Kobe, Japan}

\begin{document}

\section{Introduction}

Due to asymptotic freedom, dynamical mass gap generation and non-trivial topology, the $O(3)$ non-linear sigma model in 2D is not only an excellent toy model for Yang-Mills theory and QCD in 4D, but also an important condensed matter system. Physically the system describes a one-dimensional chain of $O(3)$ rotors. Although many analytic results with varying level of assumptions exist \cite{Zamolodchikov:1977nu,*Wiegmann:1985jt,*Dunne:2015ywa}, the system is far from well understood. The related $O(N)$ and $CP(N-1)$ systems are susceptible to large $N$ calculations, and we mention in passing that there was an interesting development in the diagrammatic Monte-Carlo study of the $O(N)$ system \cite{Buividovich:2015qba} where a bosonic system with non-trivial IR physics was successfully numerically analysed using a diagrammatic expansion.

The model has a global $O(3)$ symmetry, and if this symmetry is unbroken, the particles (or low energy excitations) are classified by the charge they carry with respect to internal symmetries. Studying the system at non-zero charges is of interest if one wants to understand the low energy dynamics of the theory. One can couple chemical potentials which can be used to control the charges and therefore the particle content of the system. In particular a chemical potential can easily be coupled to one of the $O(2)$ subgroups of $O(3)$, corresponding to one of the generators $t^a,a=1,2,3$. Singling out one of the generators\footnote{Any choice of a generator, or linear combination of them is equivalent, as it only amouns to the redefinition of the generators.} we can add a chemical potential to a conserved charge, which in the present case corresponds to the \emph{total angular momentum component of the one dimensional chain of $O(3)$ rotors}. The choice of a generator represents a choice of the axis with 
respect to which we wish to induce angular momentum.  

Recently, based on low energy symmetries and large $N$ computations, a possible phase diagram of the O(3) model at finite density was conjectured in \cite{Bruckmann:2014sla}. Since the system is gapped, the increase of chemical potential enforces condensation of charges at some critical chemical potential $\mu=\mu_c$. This value represents the  dynamically generated mass of the charged excitations. Once this happens the $O(3)$ symmetry is broken down to $O(2)$ by the presence of charges for one of the three generators (i.e., it is broken explicitly by singling out one direction of angular momentum which is populated). The low energy effective theory is therefore a two-dimensional $O(2)$ theory, which can have vortex defects. In \cite{Bruckmann:2014sla} it was argued that the effective coupling of the effective $O(2)$ model for $\mu\gtrsim \mu_c$ is strong enough so that the vortices would percolate until some $\mu=\mu_{KT}>\mu_c$ (see Section \ref{sec:results}, however), which would be signaled by a 
Kosterlitz-Thouless-Berezinskii transition, forcing vortices to combine into pairs. The vortex pairs are instantons of the O(3) model, whose size modulus is related to the distance between the vortex and the anti-vortex they are made of. The vacuum picture is that of small instantons. 

Incidentally this is very similar to what is believed to happen in QCD at asymptotically high densities: the color Cooper pairs form, and Higgs the gauge fields. This in turn, forces instantons to be small. As one reduces the chemical potential the instantons start to grow, and it was conjectured that at some critical density instantons fractionate into smaller objects \cite{Parnachev:2008fy,*Zhitnitsky:2008ha,*Zhitnitsky:2013wfa}. Although there is no complete understanding of what these instanton constituents are in four-dimensions, it is possible that they are related to instanton-monopoles which have been a topic of increasing interest in recent years, both in theoretically controllable regimes as well as in lattice field theory and phenomenology (see e.g. \cite{Poppitz:2012sw,*Bornyakov:2014esa,*Larsen:2015tso} and references therein).

Although QCD and non-linear sigma models are undoubtedly different, these similarities are striking. It is therefore of interest to study the phase diagram of the O(3) model from first principles, i.e., from lattice simulations. There is a technical difficulty here, however, generically present in systems with chemical potential: The complex action problem. Although plenty of work and attempts have been made to overcome this problem, no satisfactory solution exists to date for QCD at finite baryon density. However, an extremely elegant solution for several bosonic and some fermionic systems exists, which employs an exact rewriting of the partition function in terms of \emph{dual variables} living on links which are generically integer valued for compact symmetries. What is very important is that the dual variable fugacity is real and positive, which permits importance sampling and allows Monte Carlo simulations to be performed. The results we present here are based on such simulations. 

\section{Lattice action, dual variables and chemical potential}

Before we present the lattice results let us discuss some generalities of the system. The lattice action of the $O(3)$ model is given by
\be
S=-J\sum_{x,\nu}\bm n(x)\cdot\bm n(x+\hat\nu) \; ,
\ee
where $\bm n(x)=(n_1(x),n_2(x),n_3(x))$ is a 3-component field with unit length and where $J=1/g_0^2$ is the inverse coupling. It is clear that the transformation $\bm n(x)\rightarrow O\bm n(x), O\in O(3)$ is a global symmetry of the above actions, and due to Noether's theorem, there is a conserved charge associated with it. We use spherical coordinates such that  $\bm n(x)=(\cos\phi(x)\sin\theta(x),\sin\phi(x)\sin\theta(x),\cos\theta(x))$. We now add a chemical potential for the charge associated with the $O(2)\in O(3)$ symmetry, which changes $\phi(x)\rightarrow \phi(x)+\text{const}$. The simplest way to do this is to introduce a background gauge field for the symmetry and to set $A_\nu(x)\rightarrow i\mu \, \delta_{\nu,0}$, where $\mu$ is the chemical potential. In this way we obtain the action
\be
S=-J\sum_{x,\nu}\Big[\cos\theta(x)\cos\theta(x+\hat\nu)+\sin\theta(x)\sin\theta(x+\hat\nu)\cos(\phi(x+\hat{\nu})-\phi(x)-i\delta_{\nu,0}\mu)\Big]\;.
\ee
The action has a non-vanishing imaginary part, i.e., it suffers from the complex action problem.

However, as already mentioned, there exists an exact rewriting of the partition function, 
which uses flux variables living on links \cite{Bruckmann:2015sua}:
\begin{align}
Z=&\sum_{\{k,m,\bar m\}}
\left(\prod_{x,\nu}\frac{J^{k_{x,\nu}}}{k_{x,\nu}!}\frac{(J/2)^{|m_{x,\nu}|+2\bar m_{x,\nu}}}{(|m_{x,\nu}|+\bar m_{x,\nu})!\bar m_{x,\nu}!}\right)
e^{-\mu\sum_x m_{x,0}}\nonumber\\
&\prod_{x}I\left(\sum_{\nu}(k_{x,\nu}+k_{x-\hat\nu,\nu}),1+\sum_{\nu}\left[m_{x,\nu}+m_{x-\hat\nu,\nu}+2(\bar m_{x,\nu}+\bar m_{x-\hat\nu,\nu})\right]\right)\nonumber\\
&\prod_{x}\delta\left(\sum_{\nu}(m_{x,\nu}-m_{x-\hat\nu,\nu})\right)E\left(\sum_{\nu}(k_{x,\nu}+k_{x-\hat\nu,\nu})\right)~,
\label{dual}
\end{align}
where $m_{x,\mu}\in\mathbb{Z}$ and $\bar m_{x,\mu}, k_{x,\mu}\in\mathbb{N}_0$ are integers living on links (the dual variables). Here $\delta(n)$ 
denotes the Kronecker delta, and $E(n)$ is the evenness function which is 1 for even $n$ and vanishes for odd $n$, while 
$I(a,b)=\Gamma((a+1)/2)\Gamma((b+1)/2)/\Gamma((a+b+2)/2)$.

Notice that the chemical potential couples directly to the temporal fluxes $m_{x,0}$, and that, by taking a derivative with respect to the chemical potential, it is clear that the expectation value of these temporal fluxes are expectation values of the charge. Since the $m$-flux is conserved by the Kronecker deltas in (\ref{dual}) we can identify the temporal winding number of the $m$-worldlines with the \emph{charge of a configuration}. 

\section{The results}\label{sec:results}

\begin{figure}[tb] 
   \centering
   \includegraphics[width=0.48\textwidth]{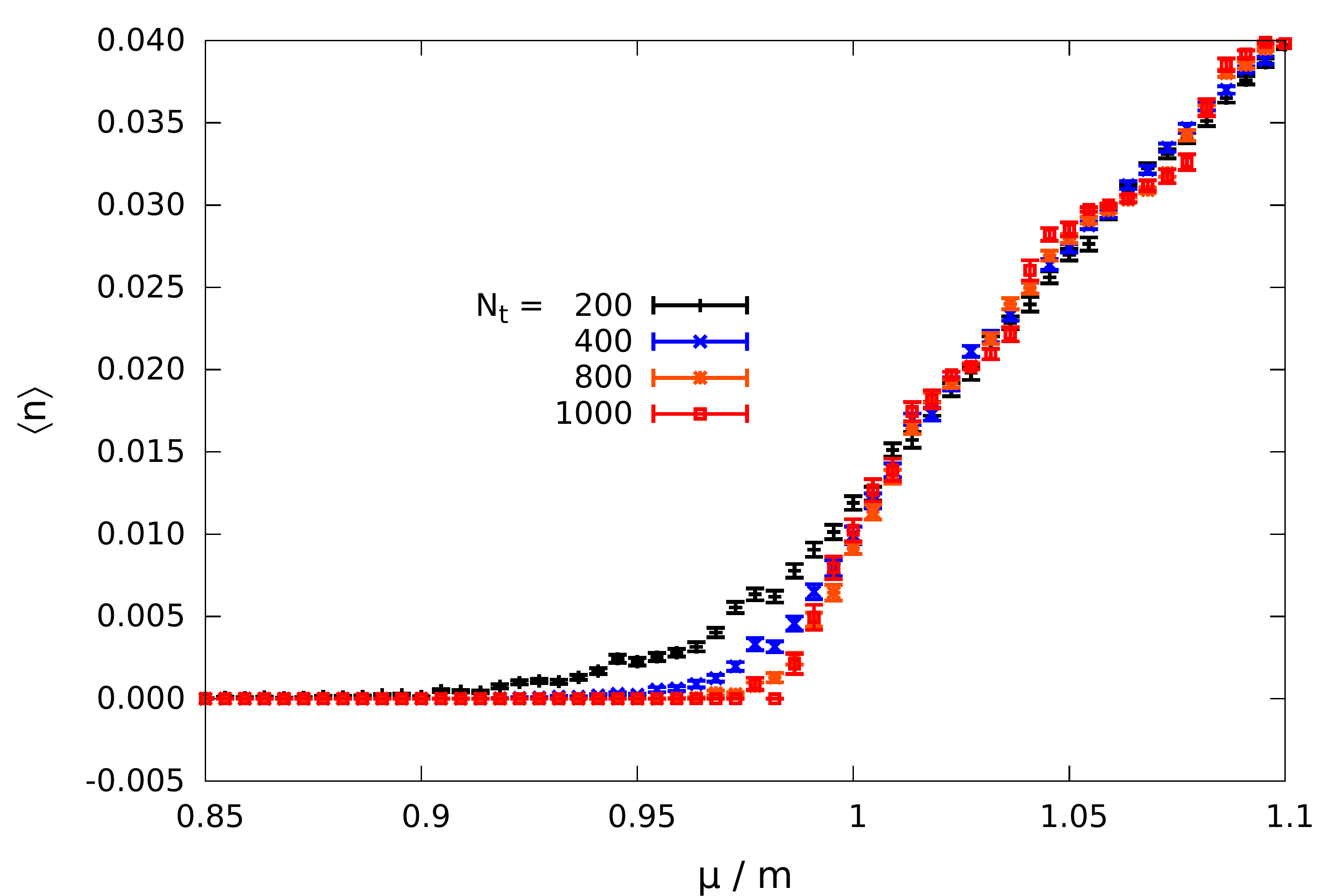}
   \includegraphics[width=0.48\textwidth]{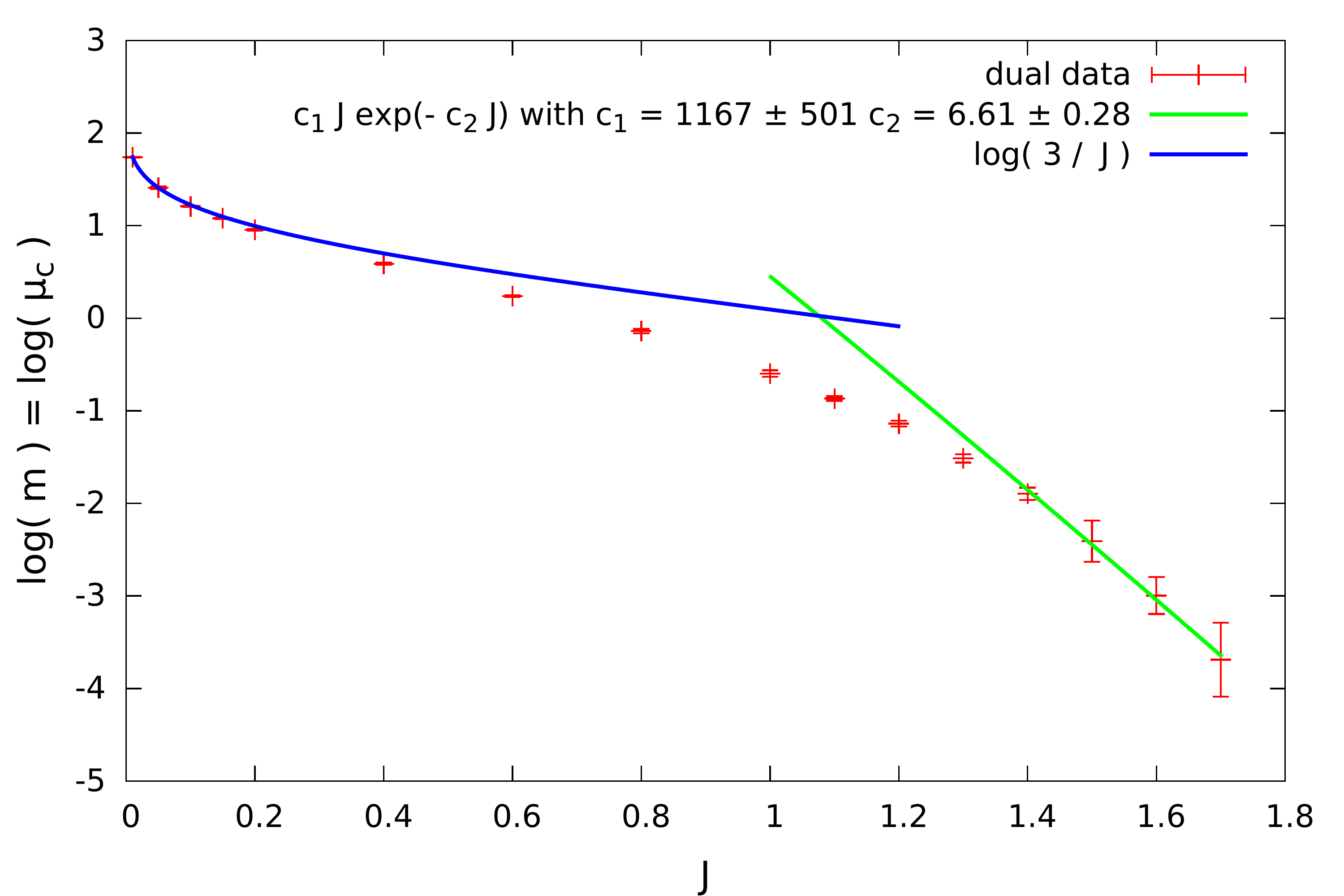} 
   \caption{Left: The charge density v.s. chemical potential $\mu$, for four temporal lattice extents $N_t$ at fixed spatial volume $N_s=100$ and coupling $J=1.3$. Right: The massgap as a function of coupling. The two solid curves are strong coupling prediction (blue) and weak coupling fit. Note that the coefficient $c_1$ is scheme dependent, while the coefficient $c_2$ is consistent with $2\pi$.}
   \label{fig:1}
\end{figure}

We have simulated the $O(3)$ model with chemical potential in the dual representation (\ref{dual}). We begin with discussing
 the charge density $\langle n \rangle = Q/L$, $Q=(\partial / \partial \mu \ln Z) / \beta$ with $L$ the spatial extent, 
 as a function of chemical potential $\mu$. We remind the reader that since the $O(3)$ model has dynamical 
 mass gap generation, it is expected that if the temperature is low enough charge condensation will set in at some 
 sharply defined non-zero value of the chemical potential, $\mu=\mu_c=m$, where $m$ is the mass of the lightest charged excitation. In $J\gg 1$ and $J\ll 1$ regimes $\mu_c$ is given by
\be
\mu_c^{J\gg 1}\propto \Lambda=C\frac{1}{a}Je^{-2\pi J} \; ,\qquad \mu_c^{J\ll 1}=\log(3/J)\;,
\label{eq:twoloop}
\ee
where $\Lambda$ is the strong scale, $a$ is the lattice spacing, and $C$ is a scheme-dependent pure number. The former comes from the fact that the strong scale is the only scale in the continuum, and the latter from an explicit calculation in strong coupling.  As can be seen from Fig.~\ref{fig:1},  this is precisely what we observe (left panel): 
At sufficiently small temperature $\langle n \rangle = 0$ for small $\mu$ and 
charge condensation sets in at a sharply defined value of $\mu$ which defines $\mu_c$. In the rhs.\ plot of Fig.~\ref{fig:1} 
the values of $\mu_c$ as a function of $J$ are compared to the above two expressions 
for the continuum- and the strong coupling limit. The data quite nicely match the continuum scaling 
for $J\gtrsim 1.4$, and the strong coupling expansion for $J\lesssim 0.4$.

\begin{figure}[tb] 
   \centering
   \includegraphics[width=.5\textwidth]{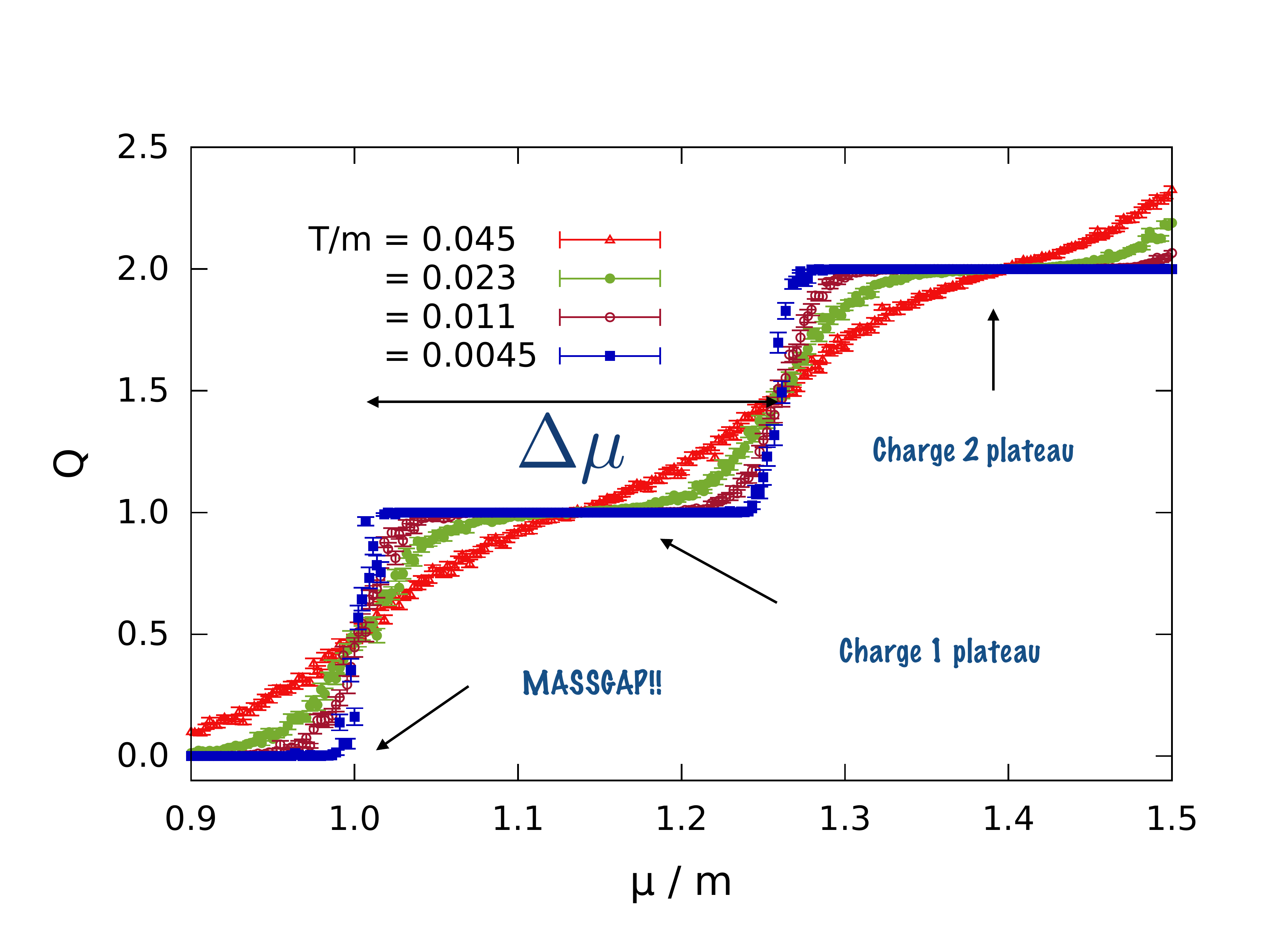}\includegraphics[width=0.5\textwidth]{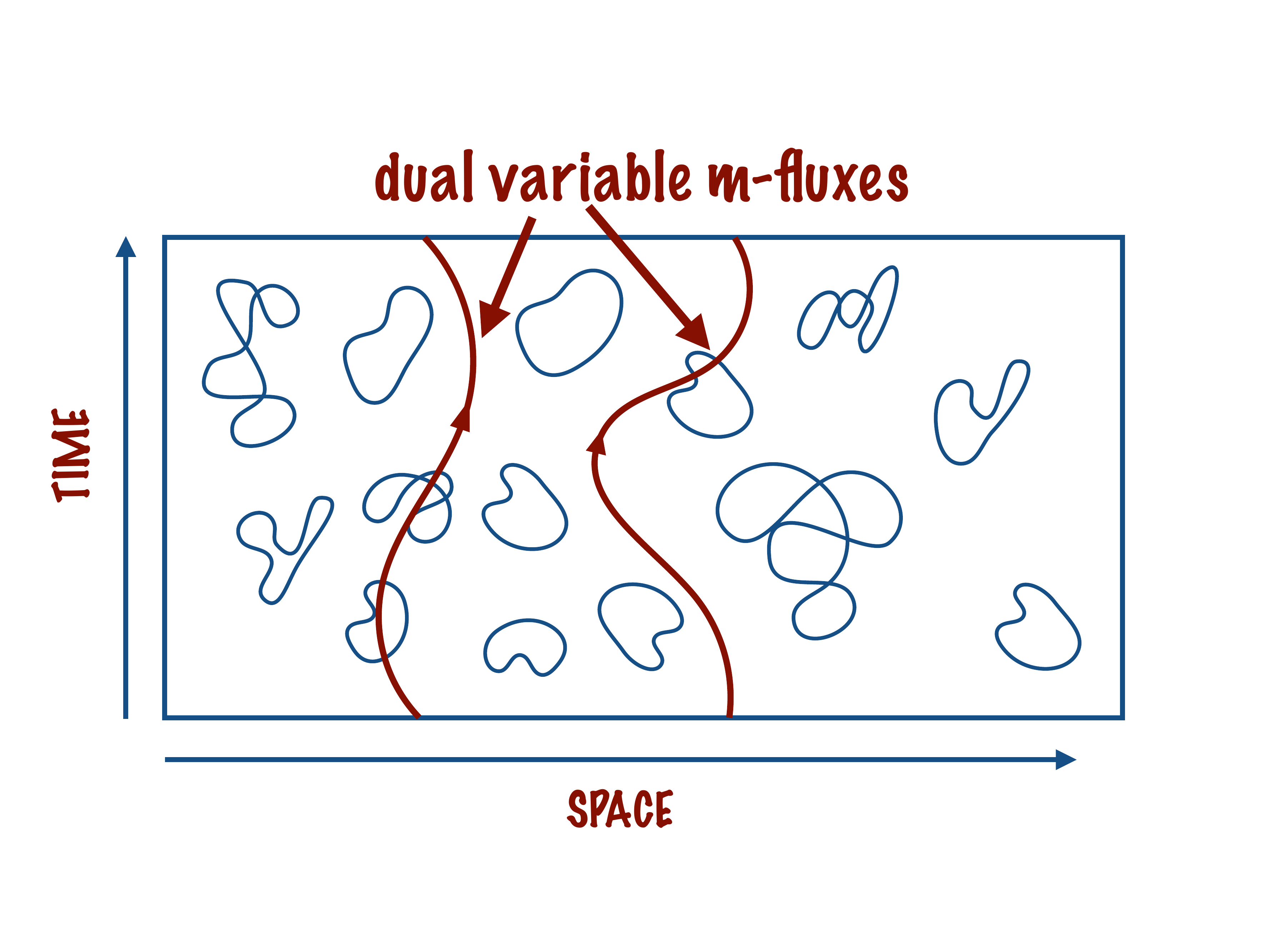}
   \caption{Left: 
   The charge $Q$ versus 
   $\mu$ at $N_s=20$, $J=1.3$ ($Lm=4.4$) for low temperatures corresponding to $N_t=100,200,400,1000$
   (from \cite{Bruckmann:2015hua}, modified). Right: a cartoon 
   of the charge-2 ensemble. The blue closed loops represent vacuum fluctuations, while the two red, winding loops represent the winding fluxes contributing to the charge of the system. These winding fluxes allow a particle worldline interpretation.}
   \label{fig:2}
\end{figure}

A feature that is not very obvious from the plot on the lhs.\ of Fig.~\ref{fig:1}, is that at sufficiently low temperature the curve is not smooth, but consists of many small steps. In the lhs.\ plot of Fig. \ref{fig:2} we show the results for a smaller volume 
at various temperatures. It is obvious that as the temperature is decreased the steps become more pronounced. The physical reason for this is that for finite spatial extent there is always a finite energy gap between the charge $Q$ and the charge $Q+1$ sectors because equal charges repel and the energy cost of placing a single particle is positive and finite. As the volume is sent to infinity this gap closes and the curve becomes continuous.
In the context of the quantum rotor model -- that has the same continuum partition function as our model 
when setting the external field to $(0,0,\mu)$ 
-- the phases between the plateaus are called 'canted' or 'unquantized ferromagnetic' \cite[Sec.\ 19.4]{SachdevQPT}. 

\begin{figure}[t] 
   \centering
   \includegraphics[width=0.48\textwidth]{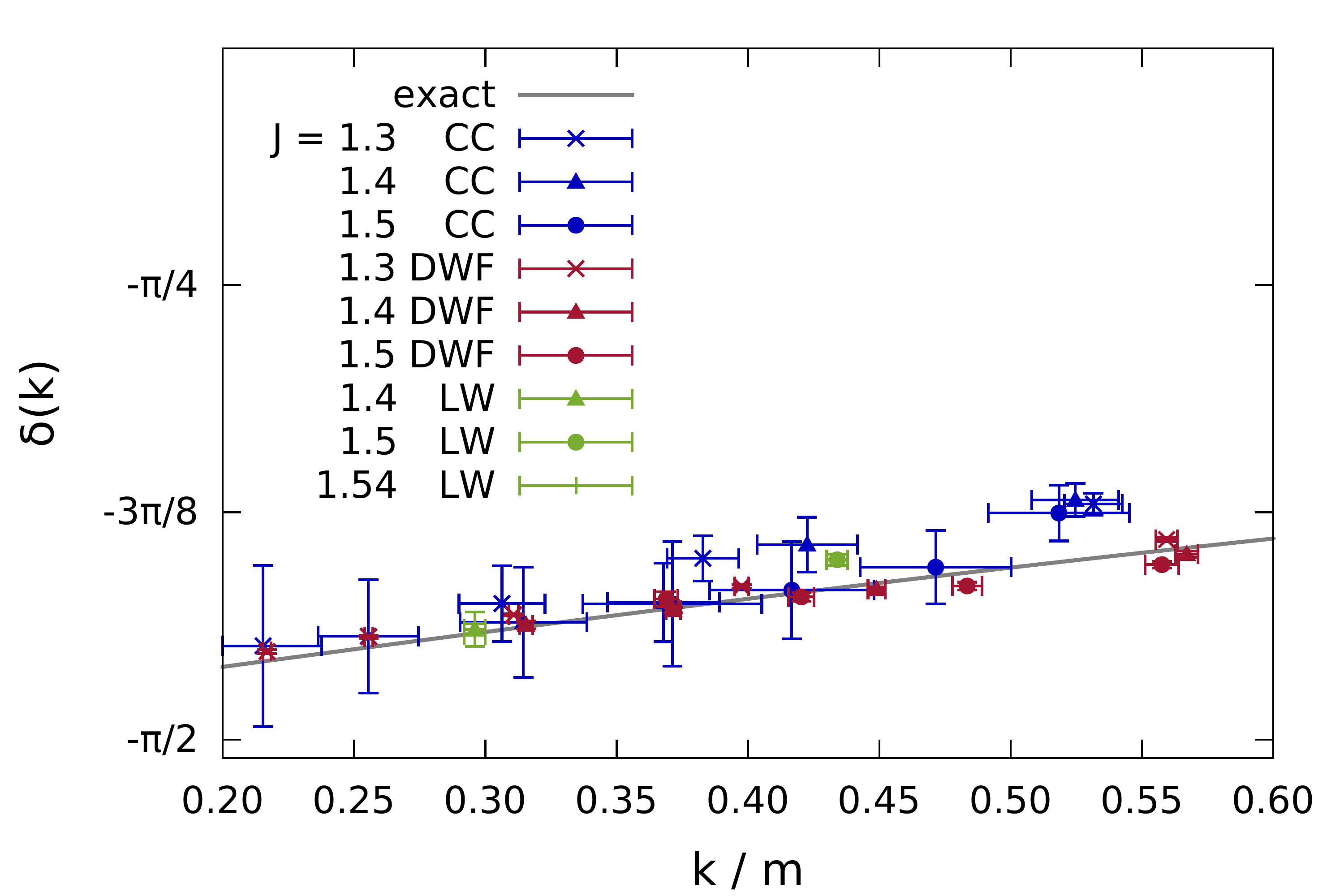}
\includegraphics[width=0.48\textwidth]{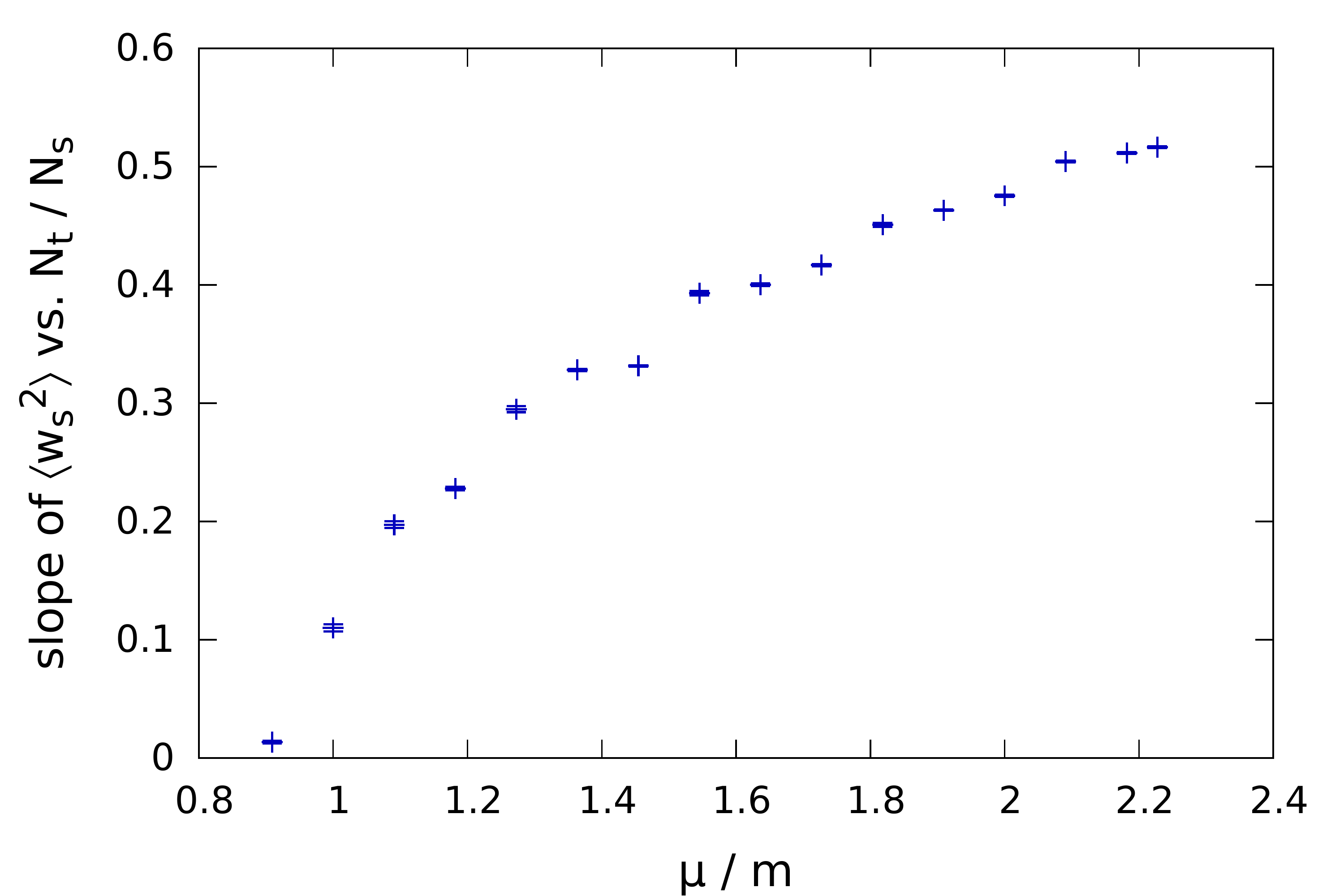}   
\caption{Left: Phase shifts obtained from the Charge Condensation Method (blue) and the Dual Wave-function method (red). 
The green symbols show data from L\"uscher and Wolff \cite{Luscher:1990ck} and the full curve the analytic solution \cite[first ref.]{Zamolodchikov:1977nu,*Wiegmann:1985jt,*Dunne:2015ywa}. 
Right: The stiffness $J_r$ as a function of chemical potential $\mu$ in mass units.}
   \label{fig:3}
\end{figure}

The small volume data are very useful for extracting information about the particle interactions. 
As was discussed in \cite{Bruckmann:2015hua}, knowing the two-particle interaction energy in 1+1 dimensions 
allows one to extract the complete scattering data, while in higher dimensions it can be used to obtain the scattering length \cite{Luscher:1990ck,*Luscher:1986pf,*Luscher:1990ux} . 
Since this interaction energy is exactly equal to the difference in $\mu$ between the onset of the charge 1 plateau and the charge 2 
plateau of the lhs.\ panel of Fig.~\ref{fig:2}, the arguments of \cite{Luscher:1990ck,*Luscher:1986pf,*Luscher:1990ux} 
can be used to obtain the scattering phase. We call this method for obtaining the scattering phases 
\emph{the charge condensation method}, as it arises from data of the charge condensation \cite{Bruckmann:2015hua}. The corresponding
results for the phase shift are shown in the lhs.~plot of Fig.~\ref{fig:3}. Note that this is a universal 
result for any system, independent of the dual variable representation, and can be used whenever the sign problem can be overcome
in a MC simulation, such as in two-flavor, two-color QCD or in QCD with isospin chemical potential. 

However, the dual flux representation has much more information about the nature of the low energy particle excitations. 
In fact the dual representation allows for the dual fluxes to be interpreted as particle worldlines (see Fig.~\ref{fig:2} rhs.). This interpretation 
can then be used to construct the wave-function of the multi-particle ground states at finite volume, by analyzing the spatial distribution 
of the worldlines. This multi-particle ground state wave-function in principle has all the information about the scattering phases and 
interaction of the particles. In the lhs.\ plot of Fig.~\ref{fig:3} we also present the results for the scattering phases extracted by this method 
which we call \emph{the Dual Wave-Function Method} (for more details see \cite{Bruckmann:2015hua}).

Let us finally address the possibility of the conjectured Kosterlitz-Thouless-Berezinskii transition. A good indicator\footnote{We would like to thank Hans Gerd Evertz for pointing this out to us.} of this transition is the so-called 
``stiffness'', which, for an effective $O(2)$ model is defined in the following way: Let $\phi(x)$ be the angular variable of the effective $O(2)$ model. In the continuum limit the kinetic Lagrangian is given by $\frac{J}{2}(\partial_\mu\phi(x))^2$. Imposing the spatial boundary conditions to be such that $\phi(x+L)=\phi_0+\phi(x)$, where $L = a N_s$ is the physical extent of the spatial direction. It can be easily checked that the classical free energy difference between the twisted and untwisted system is then simply given by $\Delta F_{class}(\phi_0)=\frac{J}{2}\phi_0^2/L$.  Quantum mechanically the coupling $J$ may be renormalized, and we can \emph{define}
\be
J_r \; \equiv \; 2L\lim_{\phi_0\rightarrow 0}\frac{1}{\phi_0^2}\Delta F(\phi_0)\;.
\ee
This quantity is usually called the \emph{stiffness}. Note, however, that the twist $\phi_0/L$ plays the role of an imaginary chemical potential in the spatial direction. It is simple to show that
\be
J_r \; = \; \frac{L}{\beta}\avg{w_s^2}\; ,
\ee
where $w_s$ is the winding number of the dual flux variable 
$m$, the one 
coupling to
$\mu$,
in the \emph{spatial} direction.
The stiffness indicates whether the system is in the gapped, vortex-percolating phase, (i.e., insensitive to the value of $\phi_0$ so that $J_r=0$), or in the gapless phase with $J_r>0$. The rhs.\ plot of  Fig. \ref{fig:3} shows our data for the stiffness from measurements of  $\avg{w_s^2}$ as a function of $\mu$.  As can be seen, the data indicates that the stiffness starts to develop a non-zero value at $\mu=\mu_c$, which seems to indicate that both, charge condensation as well as the KT transitions, occur at the same point $\mu=\mu_c$. However, one should keep in mind that the data shown is for small ratio $\beta/L$ of the inverse temperature $\beta$ and the volume $L$, and the system may have a quantum phase transition depending on the order  of limits $L\rightarrow\infty$ and $\beta\rightarrow\infty$. A more detailed analysis of this scaling with temperature and volume still needs to be performed before final conclusions can be made. 

\section{Conclusions}

We have presented several results from simulations of the O(3) non-linear sigma model at non-zero charge density in the dual variable representation. 
We have discussed two aspects of the theory: (1) The worldline interpretation of the dual fluxes and scattering phases \cite{Bruckmann:2015hua}.
(2) The phase structure as a function of chemical potential and the possibility of a KT transition. Concerning (2) we have seen that preliminary results 
seem to indicate that the KT transition happens at the same critical value of the chemical potential as the charge condensation transition. 
However, this result may be very sensitive to the order of infinite volume and zero temperature limits, and at the moment final conclusions must be 
deferred until a proper scaling analysis is performed. Concerning (1) we have demonstrated \cite{Bruckmann:2015hua} 
that the scattering phases can be reliably 
extracted from the distribution of the worldlines. This is not only an alternative approach to computing scattering data in lattice simulations, but also
reveals a deep connection between different aspects of a quantum field theory, i.e., its phenomenology in the condensed phase and scattering properties.

\bibliographystyle{JHEPm-nt}
\bibliography{lattice2015}

\end{document}